\documentclass[journal]{IEEEtran}
\ifCLASSINFOpdf
\else
\fi
\usepackage[dvips]{graphicx}
\usepackage{setspace}
\usepackage[square,comma,sort&compress,numbers]{natbib}
\usepackage{amsmath}
\usepackage{amssymb}
\DeclareMathSizes{10}{8}{7}{6}
\IEEEaftertitletext{\vspace{-2\baselineskip}}
\begin{document}
\title{Probabilistic Rateless Multiple Access for Machine-to-Machine Communication}
\author{Mahyar~Shirvanimoghaddam,~\IEEEmembership{Member,~IEEE,}
                          Yonghui~Li,~\IEEEmembership{Senior~Member,~IEEE,}
                          Mischa~Dohler,~\IEEEmembership{Fellow,~IEEE,}
                      and Branka~Vucetic,~\IEEEmembership{Fellow,~IEEE,}
\thanks{The material in this paper was presented in part at the 2014 IEEE International Symposium on Information Theory, Honolulu, HI.

M. Shirvanimoghaddam, Y. Li, and B. Vucetic are with the Center of Excellence in Telecommunications, School of Electrical and Information Engineering, The University of Sydney, Sydney, NSW 2006, Australia (e-mail: mahyar.shirvanimoghaddam@sydney.edu.au; yonghui.li@sydney.edu.au; branka.vucetic@sydney.edu.au).

M. Dohler is with King's College London (e-mail: mischa.dohler@kcl.ac.uk)}}

\maketitle
\begin{abstract}
Future machine to machine (M2M) communications need to support a massive number of devices communicating with each other with little or no human intervention.  Random access techniques were originally proposed to enable M2M multiple access, but suffer from severe congestion and access delay in an M2M system with a large number of devices. In this paper, we propose a novel multiple access scheme for M2M communications based on the capacity-approaching analog fountain code to efficiently minimize the access delay and satisfy the delay requirement for each device. This is achieved by allowing M2M devices to transmit at the same time on the same channel in an optimal probabilistic manner based on their individual delay requirements. Simulation results show that the proposed scheme achieves a near optimal rate performance and at the same time guarantees the delay requirements of the devices. We further propose a simple random access strategy and characterized the required overhead. Simulation results show the proposed approach significantly outperforms the existing random access schemes currently used in long term evolution advanced (LTE-A) standard in terms of the access delay.
\end{abstract}
\IEEEpeerreviewmaketitle
\section{Introduction}
\IEEEPARstart{M}{achine}-to-Machine communications have emerged as a promising technology to enable trillions of multi-role devices, namely machine-type communications (MTC) devices, to communicate with each other with little or no human intervention \cite{TUbiq,M2MMInternet}. It has many potential applications, such as intelligent transportation systems (ITS), healthcare monitoring, retail, banking, smart grids, home automation and so on. It is expected that in the next a few years over 2 billion MTC devices will become directly attached to cellular networks to provide M2M communications \cite{ResAlocM2M}. Thus, there will be a massive number of MTC devices with no/low mobility \cite{NovelFixed} in each cell, which is significantly more than the number of users in current cellular networks. Moreover, M2M traffic involves a large number of short-lived sessions, attempting to deliver a small amount of data (few hundred bits) to the base station, which is quite different from those in human-to-human (H2H) communications. Such differences motivate researchers around the globe to optimize the current cellular networks to effectively enable M2M communications \cite{RadioRAinLTEA,gotsis2013analytical}.
\subsection{Motivation and Related Work}
In most existing wireless access networks, the first step in establishing an air interface connection, is to perform random access (RA) in a contention manner. In fact, short-lived sessions with small amount of data in M2M communications, makes it inefficient to establish dedicated resources for data transmission \cite{M2MRA}. Such a random access approach will not work effectively when the number of nodes is very large, due to frequent transmission collisions, leading to network congestion, unexpected delays, packet loss, high energy consumption, and radio resource wastage \cite{M2MRA}. Thus, one of critical challenges in M2M communications is to design an effective medium access scheme to support a large number of devices. In current RA approaches,  when two or more devices select the same RA preamble in the first phase of the contention phase, a collision will occur and the respective devices will not be scheduled for data transmission. To reduce the access delay in M2M communications, several schemes have been proposed, such as \textit{dynamic allocation} \cite{EnhanceLTE,AutRA}, \textit{slotted access}, \textit{group-based} \cite{TUbiq,EEMacessM2M}, \textit{pull-based}, and \textit{access class barring} techniques \cite{PriorACB,ClassBarr}. Although these approaches can reduce the access collisions to a certain degree, most of them still suffer from very high access delays in highly dense networks. The main idea behind these schemes is to delay the retransmission of the access request for a random/fixed amount of time, thus increasing the access probability within a relatively short time. This is however inefficient in M2M communications due to small short burst transmissions of devices which mainly do not require the whole RB for their transmission. This means that for M2M communications with
a very large number of devices there might not be enough RBs to be orthogonally allocated to the devices, which significantly increase the access delay even if the random access requests are delivered correctly.

Additionally, different MTC devices have diverse service requirements and traffic patterns in M2M communications. Generally, we can divide the traffic types into four different categories. The first type is the alarm traffic, which is completely random and its probability is very low; however, it has a very strict delay requirement.  The second traffic type can be modeled by a random Poisson distribution with the parameters depending on the application \cite{FunThroM2M}. The regular traffic, such as smart metering applications, is the third traffic type, and the last type is the streaming, like video surveillance applications. Current proposals for enabling M2M communications, did not consider the priorities among devices and different quality of service (QoS) requirements. These approaches are mostly inefficient for M2M communications as they are generally designed for a fixed payload size and thus, cannot support M2M applications with different service requirements.

Recently, a systematic framework has been developed in \cite{FunThroM2M,PowerEfficient} to understand the fundamental limits of M2M communications in terms of power efficiency and throughput. However, they did not provide a systematic approach to develop an efficient communication protocol to approach these limits. Here, we consider a realistic model for M2M communications, which supports both regular and random traffics with different delay and service requirements. We develop a practical transmission scheme for M2M communications based on recently proposed analog fountain codes (AFCs) \cite{MahyarLetter} to enable massive number of devices communicating with a common base station (BS) while satisfying QoS requirements of all devices. We further show that the proposed scheme can closely approach the fundamental limits of M2M communications in terms of throughput and can satisfy the delay requirements of all devices at the same time. The main contributions of this paper are summarized next.
\subsection{Contributions}
\subsubsection{Efficient Random Access Proposal} We consider the contention-based random access in the proposed probabilistic multiple access for M2M communications. In the contention process, each device transmits an RA preamble to the BS to establish a communication link with the BS. In existing RA schemes in M2M communications, the devices will be identified by the BS in the contention phase if each device has chosen an RA preamble which is not selected by other devices, and a resource block (RB) will be allocated to the device for the data transmission. For a system with a large number of devices,  several devices may choose the same RA preamble and thus the BS cannot recognize the devices in the RA process. Moreover, even if the BS can identify all active devices, the number of RBs is not sufficient to support all of them. To overcome this problem, we propose a new contention mechanism. We group the RA preambles into several sub-sets according to delay requirements of MTC devices. Then, a sub-set of RA preambles is assigned for those devices which have the same delay requirement. In the contention period, each device selects a specific RA preamble based on its delay requirement from the associated subset of RA preambles. By detecting the RA preamble of the device, the BS then knows which subset this RA preamble belongs to and thus knows its delay requirement. The BS detects the number of devices which have selected the same RA preamble and broadcasts this information to the devices. The devices which have selected the same RA preamble then transmit in the same RB, and their identities will be transmitted along with their payload data and will be later recognized by the BS after the decoding. In this way, each available RB will be effectively used by a number of devices and RA collisions is handled in an efficient manner to simultaneously support collided devices.
\subsubsection{Efficient Maximum Throughput Transmission} Our second contribution in this work is to represent the multiple access process as a special kind of analog graph code. This enables us to draw on the recently proposed capacity approaching analog fountain codes \cite{MahyarLetter} to design  the optimal access protocol achieving the sum-rate capacity of multiple access channel.

Fountain codes have been used to facilitate the multiple access transmissions over erasure channels \cite{RatelessMAErasure,Mod-Sum,SigSag,SRAGraph}, where collided packets are used in an effective way to increase the overall system throughput; however, they cannot be used for wireless channels which are affected by fading and noise. In \cite{AchGausSpatCoup}, it has been shown that the capacity of a Gaussian multiple access channel can be achieved by using spatially coupled sparse graph multi-user modulation with an iterative interference cancelation scheme at the receiver. However, this approach performs decoding and interference cancelation separately and its complexity increases significantly as the number of access users increases. Thus, it will not be feasible for M2M applications with a massive number of users. To overcome this problem, we proposed a simple capacity-approaching analog fountain code (AFC) in \cite{MahyarLetter} with linear complexity in both the encoder and decoder. Thanks to the interesting linear property of AFCs, the summation of AFC coded signals of multiple simultaneous transmitted users still forms an AFC code \cite{MahyarISIT2014}. Thus the receiver can represent the multiple access transmission by a special kind of graph codes, and the BS can jointly decode them by using a standard belief propagation (BP) decoder. The proposed probabilistic multiple access AFC (MA-AFC) scheme enables the devices to transmit in a probabilistic manner in a way to form an equivalent AFC code at the BS, and approach the sum-rate capacity of the multiple access channel \cite{MahyarISIT2014}.
\subsubsection{QoS aware Random Transmission} Different devices in M2M communications have diverse traffic and service requirements. Existing random access schemes do not provide QoS guarantees in the random access transmissions. This precludes use in many practical M2M systems, where timing constraints are critical. How to alleviate access congestions in massive access transmissions while providing delay or QoS guarantees remains a significant hurdle in M2M access network design. In the proposed approach, the degree distribution functions of AFCs and access probabilities are optimized such that the delay requirement for each device is satisfied. We formulate an optimization problem based on the density evolution analysis of the AFC code to find the optimum degree and access probability for each device to satisfy their delay requirements. We further show that the proposed approach can simultaneously satisfy the delay requirement for all devices.

Throughout the paper the following notations are used. We use a boldface small letter to denote a vector and the $i^{th}$ entry of vector \textbf{v} is denoted by $v_i$. A boldface capital letter is used to denote a matrix and the $i^{th}$ element in the $j^{th}$ row of matrix \textbf{G} is denoted by $g_{i,j}$. The transpose of matrix \textbf{G} is denoted by $\textbf{G}'$. A set is denoted by a curly capital letter and the cardinality of set $\mathcal{S}$ is represented by $|\mathcal{S}|$.

The rest of the paper is organized as follows. In Section II, we present the system model. The proposed probabilistic multiple access scheme for M2M communications is presented in Section III. Section IV presents the asymptotic analysis of the belief propagation decoding of the MA-AFC scheme based on the density evolution analysis. In Section V, an optimization problem is formulated to find the optimum code parameters. Simulation results are shown in Section VI, and finally Section VII concludes the paper.
\section{System Model}
In this section, we first introduce the network and channel model in M2M communications. Then we discuss is more details the packet transmission and delay model for MTC devices.
\subsection{Network Model and Random Access}
We consider an M2M last mile access system, where $N$ MTC devices are uniformly distributed in a single cell of radius $r_0$ and communicate with a base station (BS) located at the origin. In this paper, we focus on the M2M applications where MTC devices are deployed at fixed locations, such as smart meters, sensors or cameras in buildings, roads, bridges, etc. \cite{IsRandomLTE}.

It is assumed that the channel between each device and the BS is a slow time-varying block fading channel, for which the channel remains constant within one transmission block but varies slowly from one block to the other. We consider a time division duplex (TDD)-based wireless access system, where the channel gain of the uplink is assumed to be the same as that of the downlink \cite{OpPowerAlloc}. With this assumption, each device can estimate the uplink channel gain from the pilot signal sent periodically over the downlink channel by the BS. Using the pilot signal, MTC devices also synchronize their timing to that of the BS. The BS, however, does not have the knowledge of any channel state information (CSI). This assumption is particularly relevant in M2M communications, where due to a large number of devices, it would be impractical for the BS to obtain CSI to every MTC device \cite{RayCSI}.
 \begin{figure}[!t]
\centering
\includegraphics[scale=0.37]{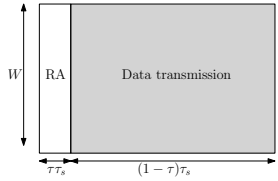}
\caption{A resource block (RB) of duration $\tau_s$ and bandwidth $W$.}
\label{RBfiglabel}
\end{figure}
The channel between each device and the BS is modeled by path loss, large scale shadowing and small scale fading effects. Thus, the received power at the BS for the signal transmitted from $\text{U}_i$ is given by:
\begin{align}
\label{pathlosformula}
P_{r,i}=P_{t,i}\gamma |h_i|^2,
\end{align}
where $P_{t,i}$ is the transmit power, $\gamma$ is the reference signal to noise ratio (SNR), defined as the average received SNR from a device transmitting at maximum power located at the cell edge, $h_i$ is the effective channel gain, defined as $|h_i|^2\triangleq \chi_i \eta_i (r_i/r_0)^{-\alpha}$, $r_i$ is the distance from U$_i$ to the BS,  $\chi_i$ is a log-normal random variable modeling shadowing gain with standard deviation $\sigma$ dB, $\eta_i$ is small scale fading modeled by exponential random variable with parameter 1, and $\alpha$ is the propagation loss factor. We further assume that the devices perform uplink power control such that the average received SNR from all devices at the BS are the same \cite{FunThroM2M} and equals to $\gamma_0$. Such an assumption has been considered for code division multiple access (CDMA) in \cite{PowerEfficient}.

Time is divided into slots of duration $\tau_s$ secs and total available bandwidth is $W$ Hz. Each time slot is referred to as a resource block (RB), where a fraction $\tau$, $0<\tau<1$, of the resource block is reserved for the random access procedure as shown in Fig. \ref{RBfiglabel}. Similar to \cite{FunThroM2M}, we assume that the number of active devices in each RB of duration $\tau_s$ is random and follows a Poisson process with rate $\lambda$. We also consider a contention-based random access strategy in M2M communications, where $N_s$ different random access preambles, which are orthogonal frequency division multiplexing (OFDM) symbols in the LTE-Advanced standard \cite{M2MRA}, are selected in a random access attempt. The BS also assign an access probability to each device according to its delay requirement. The access probabilities will be later optimized in this paper to satisfy the delay requirement of all devices. The notation used in this paper is summarized in Table \ref{NotSum} for quick reference.

It is important to note that in many M2M applications, M2M and human-to-human (H2H) communication coexists. H2H users have high priority to obtain a connection to transmit their data to the BS. Most of research work has considered that the contention-free random access is used for H2H users, that is the BS assigns a preamble and an RB to the H2H user in a timely manner \cite{PerfModelDelay}. In this work we only focus on M2M devices, where opportunistically contend for RBs through a contention-based random access, and H2H users are assumed to have access to the BS through the allocated RBs.

\subsection{Delay Model for MTC devices}
\begin{table}[t]
\caption{Notation Summary}
\label{NotSum}
\centering
\scriptsize
\begin{tabular}{|p{1cm}|p{6cm}|}
\hline
\textbf{Notation}&\textbf{Description}\\
\hline
$r_0$& Cell radius\\
\hline
$W$& Total bandwidth in Hz\\
\hline
$\tau_s$&The duration of a resource block (RB)\\
\hline
$\tau$& The ratio of the duration of RA process in an RB\\
\hline
$h_i$& The effective channel gain of device U$_i$\\
\hline
$\gamma$& The reference SNR\\
\hline
$\gamma_0$& Received SNR at the BS after power control\\
\hline
$\lambda$& Average number of active devices in an RB\\
\hline
$\lambda_i$& Packet arrival rate at the $i^{th}$ device\\
\hline
$N$& Total number of devices\\
\hline
$N_s$& Total Number of RA preambles\\
\hline
$k$& Payload size of each MTC device\\
\hline
$\textbf{b}^{(j)}$& Vector of information symbols of U$_j$\\
\hline
$\textbf{G}^{(j)}$& Generator matrix of the AFC code used in U$_j$\\
\hline
$d_c$& AFC code degree\\
\hline
$\mathcal{W}_s$& Weight set of AFC\\
\hline
$D$& AFC weight set size (AFC maximum degree)\\
\hline
$\sigma^2_w$& Average weight power of an AFC\\
\hline
$N_d$& Number of delay groups\\
\hline
$t_i$& The delay constraint of the $i^{th}$ delay group\\
\hline
$p_i$& Access probability of the $i^{th}$ device\\
\hline
$L(\textbf{s})$& Length of the random seed for devices with RA preamble \textbf{s}\\
\hline
$\epsilon$& The probability that BS fails to detect random seeds\\
\hline
$\mathbb{Z}_{k_1}^{k_2}$&$\{k_1,k_1+1,...,k_2\}\subseteq\mathbb{Z}$ for $k_1\le k_2$\\
\hline
\end{tabular}
\end{table}
\normalsize

In M2M communications, timing constraints for MTC devices typically range from 10 ms to several minutes due to the infrequent transmission features \cite{TUbiq}. For instance, some M2M multimedia applications such as video surveillance systems with strict timing constraints have the delay requirement ranging from 10 ms to 40 ms. Such a diverse QoS requirements in M2M communications intensify the need for a practical channel assignment scheme to satisfy these requirements. Moreover, different MTC devices have diverse traffic patterns; i.e., some of them are regular and some have completely random traffics. In Table \ref{DelayApp}, we divide different M2M applications based on their delay requirements and traffic patterns, according to the 3rd Generation Partnership Project (3GPP) M2M standard in \cite{TUbiq}. More specifically, we divide the MTC devices into $N_d=10$ groups based on their applications, where the maximum affordable delay in group $i$ is denoted by $t_i$.

It is important to note that MTC devices with regular traffic have predictable activities and can be completely coordinated by the BS. This means that the BS will regularly assign an RB to the devices with regular activities and they can transmit their packets through the assigned RB. This can be easily obtained through several approaches in current LTE and LTE-A standards \cite{IsRandomLTE}. However, the devices with completely random activities should contend for RBs as allocating predetermined RBs to them is not efficient \cite{TUbiq}. Therefore in this paper, we only consider MTC devices with random activities with irregular traffic patterns. This is the most challenging issue in M2M communications due to the ever increasing number of such devices.

Each active MTC device $\text{U}_i$, $i\in\mathbb{Z}_{1}^{N}$, is assumed to have $k$ information symbols to transmit within a maximum delay of $t_j$, where $1\le j\le N_d$. That is, these $k$ information symbols must be delivered to the BS within $t_j$ seconds after U$_i$ starts transmitting this message, i.e, U$_i$ belongs to the $j^{th}$ delay group. We also assume that the $i^{th}$ device has the packet arrival rate $\lambda_i$, that is in each RB, $\lambda_i$ packets are generated at the $i^{th}$ device, where each packet has to be delivered to the BS no later than $t_j$ seconds after its generation. Let us assume that $1/\lambda_i>t_j$. Therefore, if a previous packet has been delivered at the BS no later than $t_j$ seconds after its generation at the $i^{th}$ device, the new packet will be immediately sent by the device and there is no need to buffer it. On the other hand, if $1/\lambda_i<t_j$, we set the delay requirement of the device to $1/\lambda_i$ to avoid any further delay due to queuing. In other words, if the device has no memory to buffer the packets (i.e. each packet has to be successfully delivered at the BS before the generation of the next packet), the appropriate delay requirement for the device is $\min\{1/\lambda_i, t_j\}$. It is also important to note that for devices with limited buffer size, the device can change the access probability according to the number of packets in its buffer. Moreover, a packet in M2M communications has a small size (few bytes); therefore, we can assume that all the packets in the buffer of a MTC device can be transmitted in one RB \cite{PerfModelDelay}. Therefore In the rest of the paper, we only use the term delay requirement for each MTC device, and assume that the queuing delay has been taken into account when calculating the appropriate delay requirement of each device.

\begin{table}[t]
\caption{Several M2M applications with their delay constraints.}
\label{DelayApp}
\centering
\scriptsize
\begin{tabular}{|c|p{1.5cm}|p{1.5cm}|p{1.5cm}|p{1.5cm}|}
\hline
Group&Delay Constraint& Example of application&Traffic Type&Message Size\\
\hline
\hline
1&10 ms&Emergency Alarm \cite{ETSISmartGrids}&Very Unlikely &Very small\\
\hline
2&20 ms&Intelligent transport system (ITS)&Regular/Irregular&Medium\\
\hline
3&40 ms&Video streaming&Regular (Streaming)&Large\\
\hline
4&100 ms&Control messages for devices \cite{ETSISmartGrids}&Regular&Very small\\
\hline
5&500 ms&eHealth&Regular&Medium\\
\hline
6&1 s&Monitoring the Devices \cite{ETSISmartGrids}&Regular&Small\\
\hline
7&10 s&eHealth&Regular/Irregular&Medium\\
\hline
8&100 s&eHealth&Regular/Irregular&Medium\\
\hline
9&500 s&Smart meters&Regular&Medium\\
\hline
10&$>$1000 s&Smart meters&Regular&Medium\\
\hline
\end{tabular}
\end{table}
\normalsize

\section{The Proposed Rateless Probabilistic Multiple Access for M2M Communications}
In this section, we develop an efficient multiple access scheme for M2M communications based on analog fountain codes. For this aim, we first briefly introduce AFC codes. Then, we present the rateless probabilistic multiple access for M2M communications, referred to as multiple access analog fountain coding (MA-AFC).
\subsection{Analog Fountain Codes}
Analog fountain codes (AFC) were originally proposed in \cite{MahyarWCNC,MahyarLetter} as an effective adaptive transmission scheme to approach the capacity of wireless channels. In AFC,  the entire message of length $k$ binary symbols is first modulated using binary phase shift keying (BPSK) modulation to obtain $k$ modulated information symbols, $b_i\in\{-1,1\}$,  where $i\in \mathbb{Z}_{1}^{k}$. Then according to the predefined degree distribution function, $d_c$ randomly selected modulated information symbols are linearly combined with real weight coefficients to generate one coded symbol. For simplicity, we assume that the code degree, $d_c$ is fixed and weight coefficients are chosen from a finite weight set, $\mathcal{W}_s$ with $D$ positive real members, as follows:
\begin{align}
\mathcal{W}_s=\{w_i\in \mathbb{R}^{+}|i\in\mathbb{Z}_{1}^{D}\},
\end{align}
where $\mathbb{R}^{+}$ is the set of positive real numbers. Let us assume that \textbf{b} is a BPSK modulated vector of dimension $k\times 1$, and \textbf{G} is the generator matrix of dimension $m\times k$, then AFC coded symbols, \textbf{c}, are generated as follows:
\begin{align}
\label{orgAFC}
\textbf{c}=\textbf{Gb},
\end{align}
where $m$ is the number of coded symbols, only $d_c$ elements of each row of matrix \textbf{G} are nonzero, and each nonzero element of \textbf{G} is randomly chosen from the weight set $\mathcal{W}_s$. To further enhance the performance, in \cite{MahyarLetter} we proposed to use a high-rate precoder to encode the original data before applying the AFC code. To fully utilize the constellation plane, i.e., both in phase and quadrature phases, two consecutive coded symbols compose one signal as $c_i+\sqrt{-1}c_{i+1}$. We also use a standard belief propagation (BP) decoder to decode AFC codes. Further details of AFC encoding and decoding can be found in \cite{MahyarLetter,MahyarISIT2014}.

\subsection{Multiple Access AFC for M2M communications}
The proposed probabilistic multiple access AFC (MA-AFC) scheme contains two steps. In the first step, namely \emph{contention phase}, the random access requests are sent by the devices to the BS and a connection is established between the devices and the BS. In the second phase, namely \emph{data transmission phase}, the actual payload data along with the device identity (ID) is transmitted by the devices to the BS. 

\subsubsection{The contention phase (CP)}
To apply AFC codes to M2M communication systems and satisfy delay requirements of all devices, we need to slightly modify the random access procedure in LTE-A \cite{M2MRA}. The BS first partitions $N_s$  RA preambles into $N_d\le N_s$ subsets, according to delay constraints of MTC devices, and each group of RA preambles are assigned for one delay group. The BS then broadcasts this information to all devices. Let $\mathcal{S}_i$ denote the set of RA preambles allocated to the $i^{th}$ delay group, then an MTC device with delay constraint $t_i$ will randomly select an RA from $\mathcal{S}_i$ in the contention period. By detecting the RA preamble of the device, the BS will know which subset this RA preamble belongs to and thus know its delay requirement.

In the first step of CP, an MTC device with delay constraint $t_i$ randomly selects an RA preamble from set $\mathcal{S}_i$ and transmits it to the BS. Let $\textbf{s}\in \mathcal{S}_i$ denote the selected RA preamble, then, it is possible that more than one devices with the same delay constraint $t_i$ have chosen $\textbf{s}$. As we assume that all MTC devices perform power control, such that the received signals from different devices will have the same received power, $\gamma_0$, the received power at the BS for the RA preamble $\textbf{s}$ will be proportional to the number of devices which have selected RA preamble $\textbf{s}$. Let  $\mathcal{U}(\textbf{s})$ denote the set of MTC devices, which have selected the same RA preamble $\textbf{s}$. Since we assume that the RA preambles are orthogonal, the received signal corresponding to RA preamble \textbf{s} is given by:
\begin{align}
\textbf{y}=|\mathcal{U}(\textbf{s})|\sqrt{\gamma_0}~\textbf{s}+\textbf{z},
\label{RApreamble}
\end{align}
where \textbf{z} is the additive white Gaussian noise (AWGN) vector. The average received signal power at the BS can then be calculated as $|\mathcal{U}(\textbf{s})|^2\gamma_0$. Since the BS knows $\gamma_0$, it can easily estimate $|\mathcal{U}(\textbf{s})|$.

\begin{figure}[!t]
\centering
\includegraphics[scale=0.4]{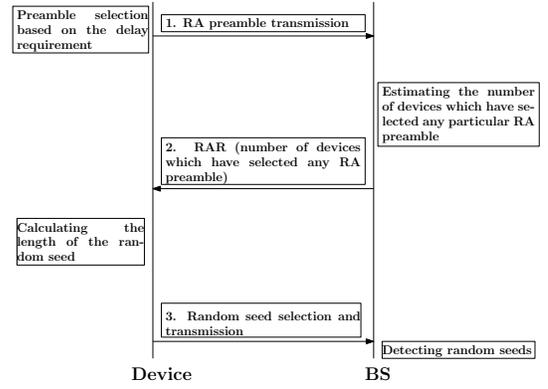}
\caption{The proposed random access procedure for M2M communications.}
\label{RAFignew}
\end{figure}

In the second step of CP, the BS transmits a random access response (RAR) message including the number of devices which have selected each RA preamble, the allocated RB for each detected RA preamble, and timing information and optimized access probabilities. In the third step of CP, each device first calculates the length of a random seed based on the received information from the BS, and then generates a random seed with the determined length and sends it to the BS. We assume that the random seeds are orthogonal sequences. Once the BS and each device have shared the same seed for the random encoding structure of AFC, they can construct the same bipartite graph to perform the AFC encoding and decoding. Fig. \ref{RAFignew} shows the proposed random access procedure for M2M communications. 

Let $L(\textbf{s})$ denote the length of the random seed selected by devices which have selected the RA preamble \textbf{s} in the first step of CP. The BS fails to uniquely detect these random seede with the probability at most $\epsilon$, which is given by \cite{FunThroM2M}:
\begin{align}
\epsilon=\left(1-\frac{1}{2^{L(\textbf{s})}-1}\right)^{|\mathcal{U}(\textbf{s})|-1}.
\end{align}
This is followed from the fact that each random seed can be detected at the BS if it has been generated by only one device. Thus, the minimum length of the random seed $L(\textbf{s})$ required to have a non-detectable probability lower than a predefined value $\epsilon>0$ is as follows:
\begin{align}
\label{ovthirdphase}
L(\textbf{s})=\left\lceil \log_2\left(1+\frac{1}{1-(1-\epsilon)^{\frac{1}{|\mathcal{U}(\textbf{s})|-1}}}\right)\right\rceil,
\end{align}
where $\lceil.\rceil$ is the ceil operator.

In the proposed RA scheme, even when two or more devices select the same preamble, they will be allocated with the same RB, thus they are transmitting in the same RB. This means that the proposed approach can support many users with a negligible collision probability, which can be controlled by the length of the random seed. Unlike the conventional RA schemes, the devices in our approach can send their random access requests whenever they have data to transmit and do not need to wait for several RA attempt to get access to the network. In fact, once the device sends a random access preamble to the BS, it will be allocated with an RB, no matter whether that preamble has been selected by other devices or not. This will significantly decrease the access delay.

\subsubsection{The data transmission phase}
After the contention phase, each active device and the BS have shared the same random seed. The devices which have selected the same RA preamble, are transmitting their AFC coded symbols in the same RB, in a probabilistic manner with the same access probability, but with different random seeds. Let $p_j$ denote the access probability assigned to U$_j$. It is important to note that devices which have selected the same RA preamble will be assigned with the same access probability. In the data transmission phase, in each time instant $\ell$, device U$_j$ generates a binary random number, $I_{j,\ell}$, which is one with probability $p_j$, i.e., $p(I_{j,\ell}=1)=1-p(I_{j,\ell}=0)=p_j$. If this random number is one, then U$_j$ generates an AFC coded symbol, $u^{(j)}_{\ell}$ and sends it to the BS. The BS also generates the same binary random number as the one in U$_j$ as they have shared the same random seed, thus it already knows in which time instants device U$_j$ is sending a coded symbol. Using this random seed, both U$_j$ and the BS can construct the same AFC code structure, so the BS is able to perform the decoding on the received coded symbols.

\begin{figure}[!t]
\centering
\includegraphics[scale=0.45]{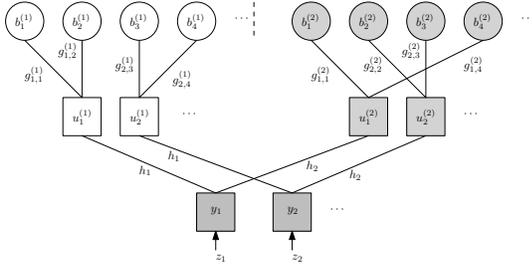}
\caption{Original AFC code graph at the BS.}
\label{graph2}
\end{figure}

\begin{figure}[!t]
\centering
\includegraphics[scale=0.45]{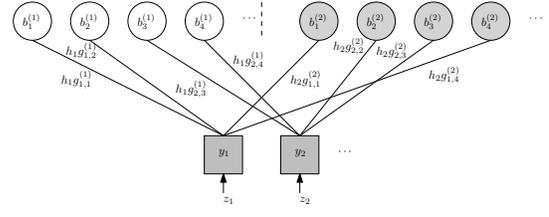}
\caption{Equivalent AFC code graph at the BS.}
\label{graph3}
\end{figure}

For simplicity, we assume that devices use the same code degree $d_c$ and the same weight set $\mathcal{W}_s$ to generate AFC coded symbols. Let $\mathcal{U}$ denote the set of devices which are currently active and transmitting AFC coded symbols to the BS based on their access probabilities in the same RB. Then, the received signal at the BS in time instant $\ell$ is given by
\begin{align}
\label{recsig}
y_{\ell}=\sum_{j\in\mathcal{U}}I_{j,\ell}~h_j~u^{(j)}_{\ell}+z_{\ell},
\end{align}
where $z_{\ell}$ is the AWGN with variance $\sigma^2_z$. As can be seen in (\ref{recsig}), the received signal at the BS is the noisy version of the sum of coded symbols of devices in $\mathcal{U}$. Since coded symbols of each device have been generated by a linear combination of modulated information symbols, the received signal at the BS can also be seen as a coded symbol of an analog fountain code, where modulated information symbols are from devices in $\mathcal{U}$. By using (\ref{orgAFC}), (\ref{recsig}) can be rewritten as follows:
\begin{align}
\label{recsig2}
y_{\ell}=\sum_{j\in\mathcal{U}}h_j~I_{j,\ell}\sum_{r=1}^{k}g^{(j)}_{\ell,r}~b^{(j)}_{r}+z_{\ell},
\end{align}
where $b^{(j)}_{r}$ is the $r^{th}$ modulated information symbol of U$_j$ and $g^{(j)}_{\ell,r}$ is the respective weight coefficient. Fig. \ref{graph2} and Fig. \ref{graph3} show the original and equivalent bipartite graphs of the AFC codes at the BS, respectively, when there are only two devices in $\mathcal{S}$ and $d_c=2$. As shown in these figures, the equivalent bipartite graph can be considered as a bipartite graph of an equivalent AFC code where the information symbols are from both users. More specifically in the first time instant where $I_{1,1}=I_{2,1}=1$, both devices transmit coded symbols to the BS. But, in the second and third time instant, only one of the devices transmits a coded symbols to the BS, i.e., $I_{1,2}=I_{2,3}=1$ and $I_{1,3}=I_{2,2}=0$ . This is because of the fact that the devices transmit coded symbols based on their access probabilities. In fact, a device U$_j$ with access probability $p_j$ transmit a coded symbols with probability $p_j$; otherwise it remains silent.

Let us define $X^{(\ell)}_{j,r}\triangleq I_{j,\ell}g^{(j)}_{\ell,r}b^{(j)}_{r}$. Then $X^{(\ell)}_{j,r}$ is a random variable with the following probability distribution:
\[p(X^{(\ell)}_{j,r}=s)=\left\{
\begin{array}{l l}
1-\frac{p_jd_c}{k}&\quad \text{if}~s=0,\\
\frac{p_jd_c}{2kD}&\quad \text{if}~s\in\mathcal{W}_s~\text{or}~-s\in\mathcal{W}_s.
\end{array}\right.\]
Thus, the mean and variance of $X^{(\ell)}_{j,r}$ can be calculated as follows:
\begin{align}
m_j&=\text{E}[X^{(\ell)}_{j,r}]=\sum_{s\in\mathcal{W}_s}\frac{p_jd_c}{2kD}(s-s)=0,\\
\sigma^2_j&=\text{E}[X^{(\ell)2}_{j,r}]=\sum_{s\in\mathcal{W}_s}\frac{p_jd_c}{kD}s^2=\frac{1}{k}p_jd_c\sigma^2_w,
\end{align}
where $\sigma^2_w=\frac{1}{D}\sum_{i=1}^{D}w_i^2$. Since in M2M communications the number of devices are very large and $X_{j,l}$'s are independent random variables, according to the central limit theorem $\sum_{j\in\mathcal{U}}\sum_{l=1}^{k}X_{j,l}$ will approximately follow a Gaussian distribution with a zero mean and the following variance:
\begin{align}
\label{AvgPY}
\sigma^2_Y=d_c\sigma^2_w\sum_{j\in\mathcal{U}}p_j|h_j|^2.
\end{align}
In other words, the received signals at the BS are normally distributed around the origin with average power $\sigma^2_Y$. As the received coded symbols at the BS can be represented as coded symbols of an equivalent AFC code, the standard BP decoding algorithm can be applied to jointly decode all active devices at the BS.

\section{Analysis of the Rateless Multiple Access for M2M Communications}
In this section, we analyze the performance of the proposed MA-AFC scheme by using a density evolution based approach. These results will be used as a basis to optimize the proposed rateless multiple access for M2M communications.

As the BS determines the allocation of RBs to the devices, we assume a general case, where the number of active devices allocated to each RB is modeled by a Poisson random variable with parameter $\lambda$. We also assume that each active device belongs to a delay group with a uniform probability. This model enables us to have a fair comparison with the throughput limit obtained in \cite{FunThroM2M}. Thus, in the rest of the paper, we only consider one RB and optimize the access probabilities for the proposed scheme to simultaneously satisfy delay requirements for all devices. .
\subsection{Optimum Coordinated Multiple Access}
Before analyzing the proposed multiple access scheme, let us first investigate the optimum coordinated multiple access, which will be used as a baseline for the proposed probabilistic multiple access AFC scheme.

We assume that the base station has already identified the devices which are transmitting in the same RB. The BS also knows all the channel information and the devices are transmitting with the same power. As shown in \cite{FunThroM2M}, the maximum common rate that can be achieved by $N$ devices transmitting in a typical RB with ordered effective channel gains $\{h_i\}_1^N$ and reference SNR $\gamma$ is given by:
\begin{align}
\mathcal{R}_c=\min_{j\in\mathbb{Z}_1^N}\frac{1}{j}\log_2\left(1+\gamma\sum_{\ell=1}^{j}|h_{\ell}|^2\right)~~ \text{bps/Hz},
\end{align}
which is simply obtained from the MAC capacity region when the transmission rate of all the devices are the same. This can be used as an upper bound on the average system common rate for the random access and later we will show how the proposed MA-AFC code can approach this limit. More specifically, when the devices perform power control in a way to have the same received SNR $\gamma_0$ at the BS, the common rate is given by:
\begin{align}
\mathcal{R}_c=\frac{1}{N}\log_2\left(1+N\gamma_0\right)~~\text{bps/Hz}.
\end{align}
Therefore, the average system throughput increases with the number of devices with a logarithmic slope. The maximum payload size $L_{opt}$ which can be successfully transmitted by each device in a resource block of duration $\tau_s$ and bandwidth $W$ is then given by:
\begin{align}
\label{optPayLoad}
L_{opt}=\frac{W\tau_s}{N}\log_2\left(1+N\gamma_0\right)~~~\text{bits},
\end{align}
which is obtained without considering the overhead for the contention phase. In fact, (\ref{optPayLoad}) provides an upper bound on the maximum payload size which can be delivered from each active device at the BS.
\subsection{Asymptotic Performance Analysis of Multiple Access AFC based on the BP Decoding}
A commonly used analytical tool for analyzing a BP decoder is density evolution \cite{ACDMAbp}, which calculates the evolutions of the message passing in the iterative decoding process. In this paper, we focus on the density evolution analysis in the asymptotic case, when the number of variable and check nodes go to infinity.

Let us refer to information symbols of device U$_i$ as Type-X$_i$ variable nodes in the bipartite graph of MA-AFC codes, for $i\in\mathbb{Z}_1^N$. Each received coded symbol at the BS may connect to modulated information symbols of various sets of devices. A simple way to represent coded symbols at the BS is to divide them into different types according to their connections to different sets of devices. For simplicity, we represent each non-empty set of devices by a vector \textbf{v} of dimension $N$, where its $i^{th}$ entry, $v_i$, is 1 if U$_i$ belongs to the set; otherwise, it is zero. Here, we refer to coded symbols which are connected to the modulated information symbols of set \textbf{v} as Type-\textbf{v} check nodes. It is easy to show that the probability that a coded symbol is of Type-\textbf{v}, $q_\textbf{v}$, is given by:
\begin{align}
q_\textbf{v}=\prod_{i=1}^{N}p_i^{v_i}(1-p_i)^{1-v_i},
\end{align}
and there are in total $2^N-1$ various types of check nodes due to the fact that the total number of nonempty subsets of devices is $2^N-1$.

Since there are $2^N-1$ types of variable and check nodes, we need to analyze the message between each specific types of variable and check nodes. In \cite{Mahyar_RC_TCOM,UEP}, an AND-OR tree analytical method was proposed to analyze the decoding error probability of LT codes \cite{Luby} with different types of variable and check nodes for erasure channels. However, the AND-OR tree approach cannot be applied directly to wireless channels. Here, we extend the conventional density evolution analysis to the case that there are multiple types of variable and check nodes.
\newtheorem{lemma}{Lemma}
\begin{lemma}
\label{GeneralDensity}
Let us consider an M2M system consisting of $N$ MTC devices wanting to simultaneously transmit their messages to a common BS. Each device U$_i$ is assigned with an access probability $p_i$ and code degree $d_i$. Let $L^{(t)}_{\text{X}_i\rightarrow \textbf{v}}$ denote the message passed from a Type-X$_i$ variable node to a Type-\textbf{v} check node in the $t^{th}$ iteration of the BP decoding algorithm at the BS. Then $L^{(t)}_{\text{X}_i\rightarrow \textbf{v}}$ can be approximated by a normal distribution with mean $m^{(t)}_{i}$ and variance $2m^{(t)}_{i}$, where $m^{(t)}_{i}$ can be calculated as follows:
\begin{align}
\label{MAAFCM}
m^{(t)}_{i}=\gamma |h_i|^2\sigma^2_wd_{v_i}\sum_{\substack{\textbf{v}\\ v_i=1}}\frac{2}{1+\sigma^2_\textbf{v}}q_\textbf{v},
\end{align}
where $d_{v_i}=md_i/k$, $\sigma^2_w=\frac{1}{D}\sum_{i=1}^{D}w_i^2$,
\begin{align}
\label{MAAFCSigma}
\sigma^2_\textbf{v}=\sum_{i=1}^{N}\gamma |h_i|^2d_iv_i\sigma^2_w \Omega\left(m^{(t)}_{i}\right),
\end{align}
and $\Omega(x)=\frac{1}{\sqrt{2\pi}}\int_{-\infty}^{+\infty}\left(1-\tanh(x-y\sqrt{x})\right)e^{-\frac{y^2}{2}}dy$.
\end{lemma}
The proof of this lemma is provided in Appendix \ref{ProfLemma1}. The BER for the $i^{th}$ device after $t$ iterations of the BP decoder, denoted by $P_{e,i}^{(t)}$, can then be calculated as follows:
\begin{align}
P_{e,i}^{(t)}=Q\left(\sqrt{m_i^{(t)}}\right),
\label{BERformula}
\end{align}
where $Q(x)=\frac{1}{\sqrt{2\pi}}\int_x^\infty e^{-z^2/2}dz$.

\subsection{Discussions on the asymptotic MA-AFC decoding performance at high SNR}
As stated before, in MA-AFC, the BS receives the sum of the coded symbols generated from active users. The BS thus tries to decode each device's data from the received signals by performing the BP decoding. Now let us consider the asymptotic performance of MA-AFC as SNR goes to infinity, thus ignoring the noise. The decoding problem in this case is equivalent to solving a system of binary linear equations, where the variables are the information symbols of all active devices. More specifically, the BS requires to solve the following system of binary linear equations to find $b^{(j)}_{l}$ for $j\in\mathbb{Z}_1^N$ and $l\in\mathbb{Z}_1^k$, given the value of $y_i$ and matrices $\textbf{G}^{(j)}$:
\begin{align}
\label{recsig2noiseless}
y_i=\sum_{j=1}^{N}\sum_{l=1}^{k}I_jg^{(j)}_{il}b^{(j)}_{l}.
\end{align}

\begin{figure}[!t]
\centering
\includegraphics[scale=0.3]{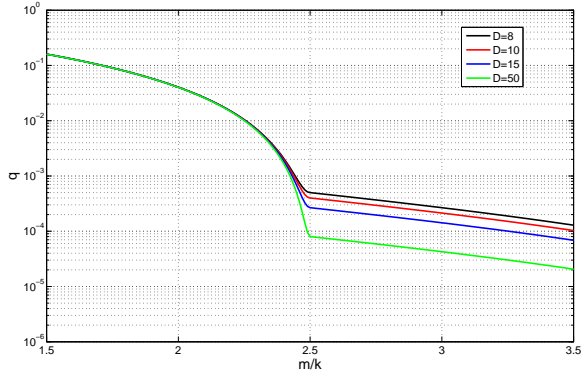}
\caption{The probability that at least two columns of the generator matrix at BS are the same versus the ratio of the number of coded symbols and that of information symbols for different weight set sizes. Each device has the access probability of 0.1 and code degree of 4.}
\label{noiselessfig}
\end{figure}

Let us define matrix $\textbf{G}\triangleq[\textbf{G}^{(1)}|\textbf{G}^{(2)}|...|\textbf{G}^{(N)}]$ and binary vector $\textbf{b}\triangleq[\textbf{b}^{(1)}|\textbf{b}^{(2)}|...|\textbf{b}^{(N)}]$. It is clear that \textbf{G} is the generator matrix of the equivalent AFC code at the BS and we have $\textbf{Y}=\textbf{G}\textbf{b}'.$ Since we consider that different devices select the weight coefficients from the same set $\mathcal{W}_s$, it is possible that two different devices select the same weight coefficients for their transmitted symbols at the same time. More specifically, at time instant $i$, the probability that coded symbols of device U$_\ell$ and U$_m$ have at least one common weight coefficient, denoted by $p_{\ell,m}$, is given by
\begin{align}
p_{\ell,m}=1-\dbinom{D-d_{\ell}}{d_m}/\dbinom{D}{d_m},
\end{align}
where $D$ is the weight set size and we assume that $D\gg d_{\ell}$ and $D\gg d_m$. It is clear that when at least two columns of \textbf{G} are exactly the same, then (\ref{recsig2noiseless}) does not have a unique solution. The following lemma gives the probability that at least two column of matrix \textbf{G} are exactly the same.
\begin{lemma}
\label{ColumnLemma}
Let us consider that $m$ coded symbols are received at the BS and each information symbol of each device is connected to at least one coded symbol at the destination, i.e., the set of non-zero elements in each column of \textbf{G} is not empty. Moreover, we assume that different devices have the same access probability $p$ and code degree $d_c$. Let $q$ denote the probability that two columns of \textbf{G} are exactly the same, then $q$ can be calculated as follows:
\begin{align}
q=\frac{\frac{(1+\beta-d_v)^2(m-d_v+1)}{Dd_v}+(d_v-\alpha)^2}{D^{d_{v}-1}\dbinom{m}{d_v-1}},
\end{align}
where $\beta=mpd_c/k$ and $d_{v}=\lceil \beta\rceil$. 
\end{lemma}
The proof of this lemma is provided in Appendix \ref{proofColumnLemma}. From this lemma, as can be also seen in Fig. \ref{noiselessfig}, even when the weight set size is small, we can always uniquely decode all devices' information symbols by transmitting more coded symbols. Moreover, when $D$ is small, devices are required to transmit more coded symbols to guarantee that the BS can uniquely decode all devices information symbols, which clearly decrease the overall system throughput. To achieve higher throughput one may consider larger code degree $d_c$, which on the other side significantly increases the decoding complexity at the BS that exponentially increases with $d_c$.

\subsection{How much overhead is required in the contention phase?}
In the first phase of the contention phase, each active device sends a RA preamble to the BS. Let $L_{cp,1}$ denote the length of the RA preamble. In the second phase, the BS sends the information about the number of devices  selected a particular RA preamble. The average number of devices which have selected the same RA preamble is $N/N_s$, where $N$ is the number of active devices and $N_s$ is the number of RA preambles. These information can be broadcasted by the BS by using at least $L_{cp,2}=N_s\lceil\log_2\left(N/N_s\right)\rceil$ information bits.  In the third phase of the contention phase, the users will select a random seed of length $L(\textbf{s})$ which is calculated by (\ref{ovthirdphase}), for a given value of $\epsilon$ and a given RA preamble \textbf{s}. Thus, at most $L_{cp,3}=N_sL(\textbf{s})$ information bits have to be sent to the BS in the third phase of the contention period. Therefore, the total overhead can be calculated as follows:
\begin{align}
\nonumber L_{ov}&=L_{cp,1}+L_{cp,2}+L_{cp,3}\\
\nonumber &=L_{cp,1}+N_s\left\lceil\log_2\left(N/N_s\right)\right\rceil\\
&+N_s\left\lceil \log_2\left(1+\frac{1}{1-(1-\epsilon)^{\frac{1}{(N/N_s)-1}}}\right)\right\rceil
\end{align}

As the BS allocates RBs to devices and devices with the same RA preamble will transmit at the same RB, each device then knows at which RB it should transmit its message based on its delay requirement. In this case, the amount of overhead can be further reduced to
\begin{align}
\nonumber L_{ov}&=L_{cp,1}+\left\lceil\log_2\left(N/N_s\right)\right\rceil\\
&+\left\lceil \log_2\left(1+\frac{1}{1-(1-\epsilon)^{\frac{1}{(N/N_s)-1}}}\right)\right\rceil
\label{overheadFinal}
\end{align}

Let $k$ denote the message length of each device and $\mathcal{R}_c$ denote the achievable common rate for a given arrival rate $\lambda$ using the proposed probabilistic multiple access approach. Then, this message can be successfully transmitted to the BS within a resource block of duration $\tau_s$ and bandwidth $W$ if the following condition is satisfied:
\begin{align}
k+L_{ov}\le \tau_sW\mathcal{R}_c.
\end{align}
In other words, the maximum message size (in bits) which can be supported by the proposed approach in an RB can be approximated by:
\begin{align}
k_{max}=\max\{\tau_sW\mathcal{R}_c-L_{ov},0\}.
\end{align}

\section{Optimization of Access Probabilities With delay Guarantees}
In this section, we formulate an optimization problem to simultaneously satisfy the delay requirements of all the devices. The access probability and code degree for each device will be optimized in a way that delay constraints of all devices are satisfied. Here we minimize the average degree of the received signal at the BS to further reduce the complexity of the decoder. Therefore, the BS should perform the following optimization process:
\begin{align}
\nonumber \min\sum_{i=1}^{N}p_id_i
\end{align}
subject to (a). $Q\left(\sqrt{m_i^{(\infty)}}\right)<\delta$, (b). $1\le d_i\le D$, and (c). $0<p_i\le1,~~\text{for}~~i\in \mathbb{Z}_i^N$, where $\delta>0$ is the target bit error rate, $m_i^{(\infty)}=\displaystyle\lim_{t\to \infty}m_i^{(t)}$, and
\begin{align}
\nonumber m_i^{(t)}=\frac{\gamma |h_i|^2\sigma^2_wd_it_i}{k}\sum_{\substack{\textbf{v}\\v_i=1}}\frac{2}{1+\sum_{j=1}^{N}\gamma |h_j|^2d_jv_j\sigma^2_w \Omega\left(m^{(t-1)}_{j}\right)}q_\textbf{v}.
\end{align}

\begin{figure}[!t]
\centering
\includegraphics[scale=0.3]{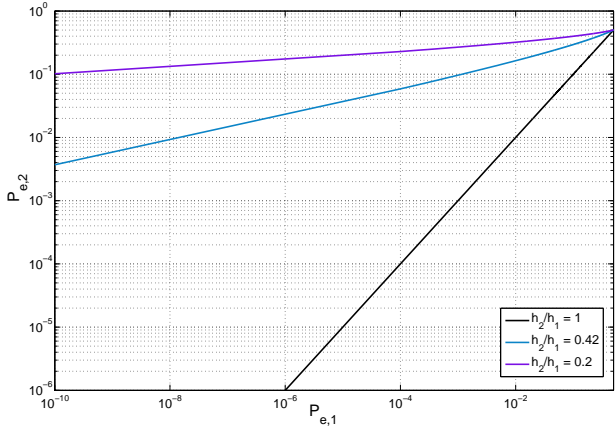}
\caption{BER of device U$_2$ versus the BER of device U$_1$ for the case that both the users have the complete access to the channel and $d_1=d_2$.}
\label{pe2pe1fig}
\end{figure}

It is important to note that $\Omega(x)$ is very close to zero for large values of $x$, so we can approximate (\ref{MAAFCM}) as follows:
\begin{align}
\label{appm}
m^{(t)}_{i}\approx \frac{2\gamma |h_i|^2\sigma^2_wd_{i}t_ip_i}{k},
\end{align}
for large $x$ values. This means that the device U$_m$ which has the highest value of  $|h_m|^2d_{m}p_m$, will be decoded sooner than other devices at the BS. Let $P_{e,i}$ denote the bit error rate of device U$_i$ at the BS. Thus, by using (\ref{appm}) we have:
\begin{align}
\label{mapprox}
\frac{m_i^{(t)}}{m_j^{(t)}}\approx\frac{|h_i|^2d_ip_i}{|h_j|^2d_jp_j},
\end{align}
where we assume that $\sigma^2_{\textbf{v}}$ is constant which is valid when the number of devices is very large. Therefore, according to (\ref{BERformula}) we have:
\begin{align}
\label{pe1pp}
P^{(t)}_{e,i}\approx Q\left(\sqrt{\left[Q^{-1}\left(P^{(t)}_{e,j}\right)\right]^2\frac{|h_i|^2d_ip_i}{|h_j|^2d_jp_j}}\right).
\end{align}
More specifically, when the devices perform power control to have the same received SNR at the BS and use the same code degree $d_c$, then (\ref{pe1pp}) can be further simplified as follows:
\begin{align}
\label{simplifyBER}
P^{(t)}_{e,i}\approx Q\left(\sqrt{\left[Q^{-1}\left(P^{(t)}_{e,j}\right)\right]^2\frac{p_i}{p_j}}\right),
\end{align}
which clearly shows that the device which has a higher access probability is decoded sooner than those with lower access probabilities. Fig. \ref{pe2pe1fig} shows the BER for device U$_2$ versus the BER of device U$_1$ for the case that both users have complete access to the channel, i.e., $p_1=p_2=1$. As can be seen in this figure, when the channel gains are the same, the BER performance is the same. Moreover, when the ratio of the channel gain of U$_1$ and that of U$_2$ decreases, the BER of U$_2$ also decreases.

By using (\ref{simplifyBER}), we can simplify the optimization problem as follows, assuming that all the devices use the same code degree $d_c$ and perform the power control to have the same received SNR $\gamma_0$ at the BS.
\begin{align}
\nonumber \min\sum_{i=1}^{N}p_i
\end{align}
subject to (a). $p_i\ge\frac{k[Q^{-1}(\delta)]^2}{2\gamma_0d_ct_i}$, and (b). $0<p_i\le1,~~\text{for}~~i\in \mathbb{Z}_i^N$. Therefore, the optimized access probabilities can be found as follows:
\begin{align}
\label{pioptfor}
p_{i,opt}=\min\left\{\frac{k[Q^{-1}(\delta)]^2}{2d_c \gamma_0 t_i},1\right\}.
\end{align}
According to (\ref{pioptfor}) we have $p_{i,opt}\propto1/t_i$. Thus, if device U$_i$ has the access probability of $p_i$, then the access probability for U$_j$ will be simply $p_j=t_i/t_j$, which further simplifies the optimization process at the BS. Furthermore, as the devices which have selected the same RA preamble will transmit in the same RB with the same access probability, the BS does not need to calculate the optimal access probability for every device. Instead, it calculated the access probabilities for each group of devices which have selected the same RB, and allocate the same access probability to devices within the same delay group.


\section{Simulation Results}
\begin{figure}[!t]
\centering
\includegraphics[scale=0.5]{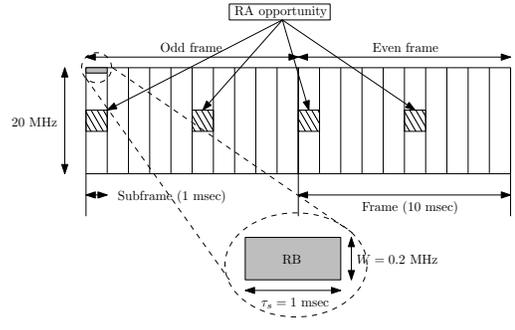}
\caption{Physical layer configuration in LTE-A \cite{IsRandomLTE}.}
\label{timeframeFig}
\end{figure}

In this section, we investigate the performance of the proposed MA-AFC scheme in M2M communication systems. We first compare our proposed MA-AFC approach with the conventional massive RA technique in the LTE-A standard \cite{ChallM2MAccess}. Like \cite{IsRandomLTE}, as shown in Fig. \ref{timeframeFig} we assume that time is divided into time frames of length 10 msec and the total system bandwidth is $20$ MHz. Each frame is further divided into 10 subframes of length 1 msec. A subframe consists of 100 RBs, each with bandwidth $W=0.2$ MHz. We also assume a total number of $N_s=60$ preambles and for simplicity we assume several configurations of RAs, which includes 1 RA attempt per two time frames, 1 RA access attempt per time frame, and 2 RA attempts per time frame. Each successful device will then be allocated with 16 RBs for the data transmission in next subframes. We assume that the packet length of devices is small (few bytes), thus a device can successfully transmit its packet within one time frame. The conventional random access approach, which is currently used in the LTE-A standard uses backoff approach \cite{M2MRA}. In this approach, when a random access attempt is unsuccessful, the device will wait for a random time before transmitting the new access request. More specifically, before the $i^{th}$ random access request, the device will wait for time $t^{(i)}_{ra}$, where $t^{(i)}_{ra}$ is randomly drawn from range $[0,X_i-1]$. The device can send a maximum number of $N_{ra}$ RA request using this backoff approach. More specifically, we assume maximum number of RA attempts $N_{ra}=10$ and backoff indexes $X_i\in\{10,20,30,40,60,80,120,160,240,320,480,960\}$ \cite{IsRandomLTE}.

\begin{figure}[!t]
\centering
\includegraphics[scale=0.3]{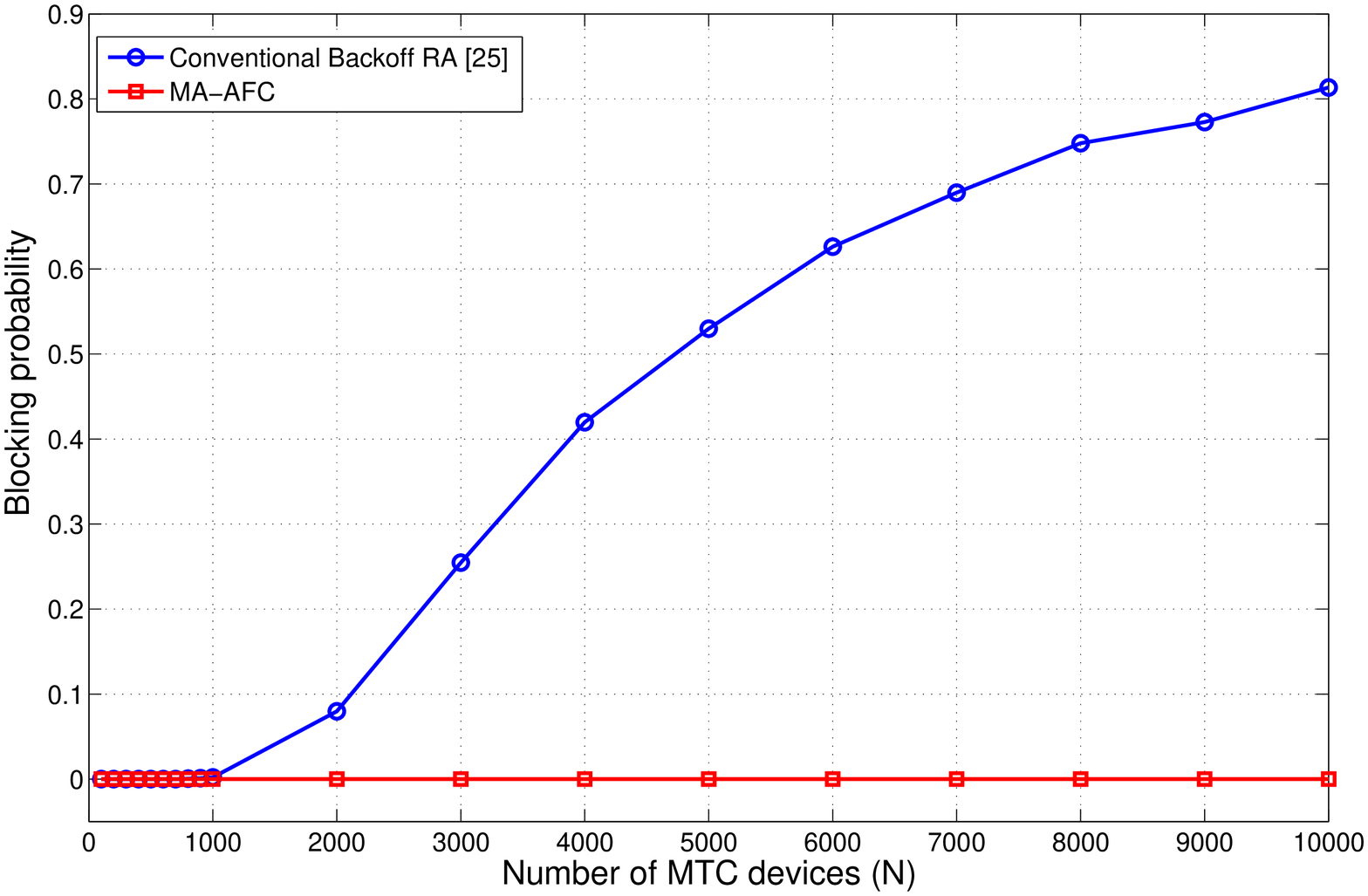}
\caption{The blocking probability for MTC devices versus the number of devices. The total number of preambles is $N_s=60$. All device in the MA-AFC case use the same code degree 8.}
\label{blovkingprobabilitfig}
\end{figure}

For the comparison, we use two metrics, blocking probability and average access delay. The blocking probability is defined as the ratio of the MTC devices failing to access the BS due to exceeding their maximum allowed random access attempts to the total number of MTC devices. The average access delay is defined as the average time elapsed from the time instant when a device sends the first access request until that it succeeds. Fig. \ref{blovkingprobabilitfig} shows the blocking probability for both the proposed and conventional approaches. As can be seen in this figure, the blocking probability for the conventional approach is very high especially when the number of devices is large. However, devices in the proposed scheme can access to the BS with a very small blocking probability, due to the fact that the devices can transmit simultaneously and they do not need to wait until they select separate preambles.

\begin{figure}[!t]
\centering
\includegraphics[scale=0.3]{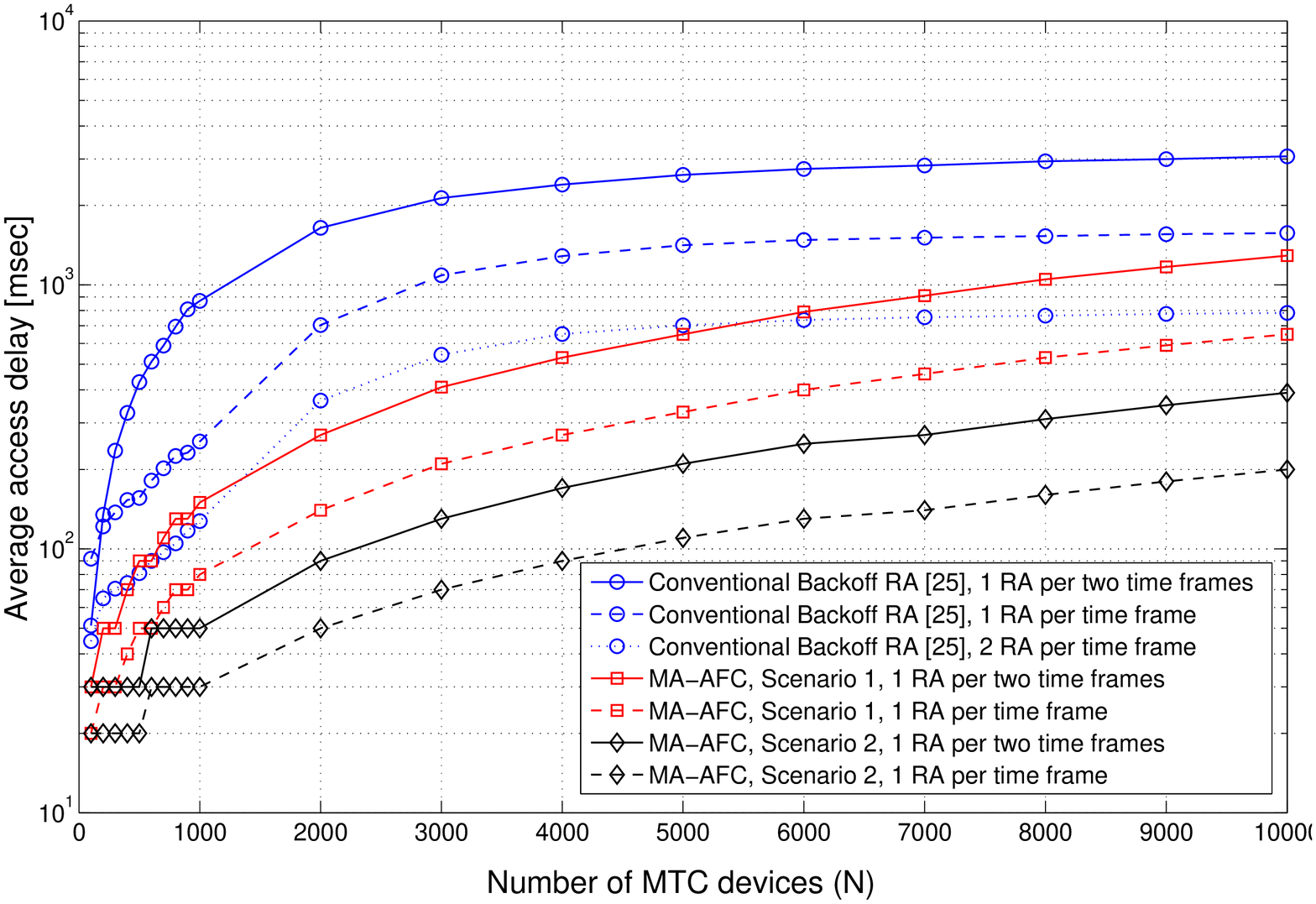}
\caption{The average access delay for MTC devices versus the number of devices. The total number of preambles is $N_s=60$. The access probability for MTC devices in Scenario 1 is $N_s/N$ and in Scenario 2 is $10N_s/N$. All device in the MA-AFC case use the same code degree 8.}
\label{Accessdelayfignew}
\end{figure}

Fig. \ref{Accessdelayfignew} shows the average access delay versus the number of devices. As can be seen in this figure, the proposed approach outperforms the conventional approach and have roughly $70\%$ less access delay. In Fig. \ref{Accessdelayfignew}, we considered two scenarios for MA-AFC. The access probability for all devices in Scenario 1 and Scenario 2 is set to $N_s/N$ and $10N_s/N$, respectively. It is important to note that the proposed approach in Scenario 2 has a lower access delay due to the fact that more devices can simultaneously transmit but it has higher complexity at the BS due to larger degrees of equivalent AFC code at the BS.

We then simulate our proposed approach within one RB, and compare the results with the fundamental throughput limits in \cite{FunThroM2M}. For this aim, we assume that the reference SNR is $\gamma = 0$ dB. This is equivalent to the case that a device transmits with 10 dBm (10mW) power over 1MHz bandwidth, a noise power spectral density of -174dBm/Hz, a receiver noise figure of 5dB, a receiver antenna gain of 14dB, a 3.76 pathloss exponent, a 128dB pathloss intercept at 1000m, and a cell radius of 1360m \cite{FunThroM2M}. We also assume that $\tau_s=1$ sec.

\begin{figure}[!t]
\centering
\includegraphics[scale=0.3]{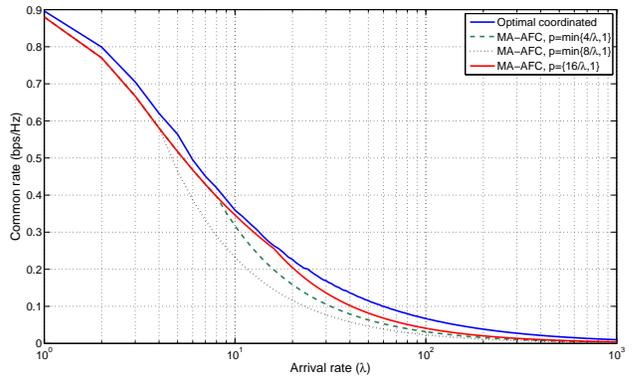}
\caption{Maximum achievable common rate versus the arrival rate for the case that the devices perform power control to have the same received SNR of 0 dB at the BS.}
\label{OprCommRate}
\end{figure}

Let us first investigate the maximum common rate which can be achieved by the proposed probabilistic MA-AFC approach in the contention-free case. Fig. \ref{OprCommRate} shows the maximum common rate versus the arrival rate for the case that the devices perform the power control to have the same received SNR of 0 dB at the BS. We consider three cases for the MA-AFC approach with different access probabilities. As can be seen in this figure, the achievable rate of the proposed approach is very close to the optimal coordinated multiple access \cite{FunThroM2M}. It is important to note that in the proposed approach we design the access probabilities in a way that the number of devices which simultaneously transmits is 4, 6, and 8, respectively, in order to minimize the decoding complexity at the BS. Fig. \ref{OprCommRate} shows that even with a small access probability, the achievable common rate of the proposed approach is very close to the optimal coordinated approach, especially when the arrival rate is very large. 

\begin{figure}[!t]
\centering
\includegraphics[scale=0.33]{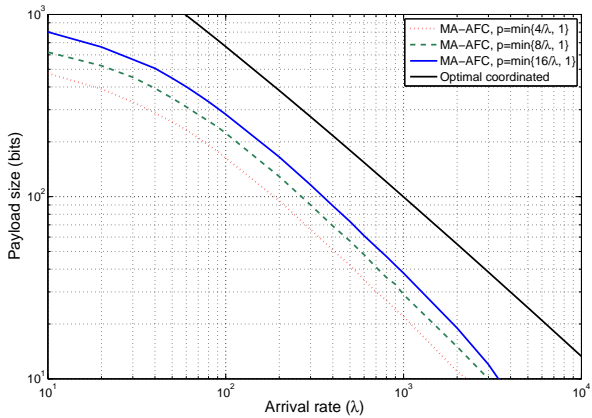}
\caption{Maximum supported payload size versus the arrival rate. Devices perform power control to have the same received SNR of 0 dB at the BS. The devices use the same code degree of 8 and the same weight set $\mathcal{W}_s$.}
\label{OptArrPayload}
\end{figure}

In Fig. \ref{OptArrPayload}, we show the maximum payload size which can be supported by the proposed approach in a resource block of duration $\tau_s=1$ sec and bandwidth $W=10$ kHz for different arrival rates. An upper bound on the maximum payload size which has been obtained from (\ref{optPayLoad}) is also shown in this figure. As can be seen in this figure, as the arrival rate increases, the payload size which can be supported decreases. This is because that the common rate is a decreasing function of the arrival rate (Fig. \ref{OprCommRate}). Moreover, as the arrival rate increases, the overhead calculated in (\ref{overheadFinal}) increases, which reduces the supported payload size. Also, by increasing the access probability, a higher common rate can be achieved, which leads to higher supported payload sizes.

\begin{figure}[!t]
\centering
\includegraphics[scale=0.29]{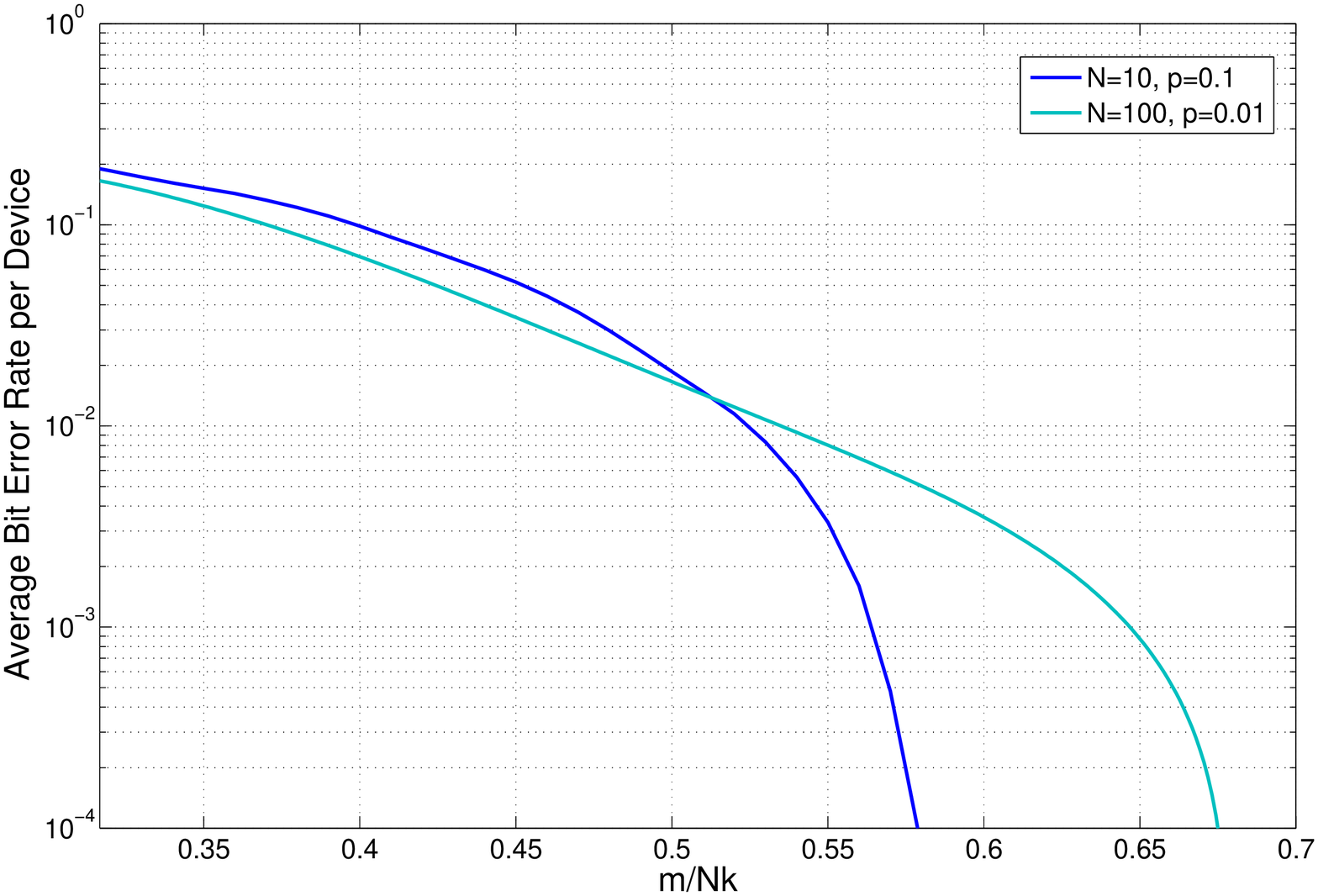}
\caption{BER versus the ratio of the number of coded symbols and the total number of information symbols. $m$ is the number of coded symbols, $N$ is the number of devices, and $k$ is the length of the message for each device. }
\label{BERSymFig}
\end{figure}

Fig. \ref{BERSymFig} shows the average BER versus the number of received coded symbols at the BS when the code degree for all devices is $d_c=8$ and the reference SNR is $\gamma=30$ dB. As can be seen in this figure, by increasing the number of devices, the BS requires more coded symbols, in order to decode all devices' information symbols. Moreover, by increasing the number of devices, the average number of required coded symbols per device remains relatively constant to achieve a certain BER. This indicates that the proposed scheme can provide a certain level of BER at the BS for each device even with a large number of devices and the decoding performance is not significantly degraded by increasing the number of devices.

\begin{figure}[!t]
\centering
\includegraphics[scale=0.3]{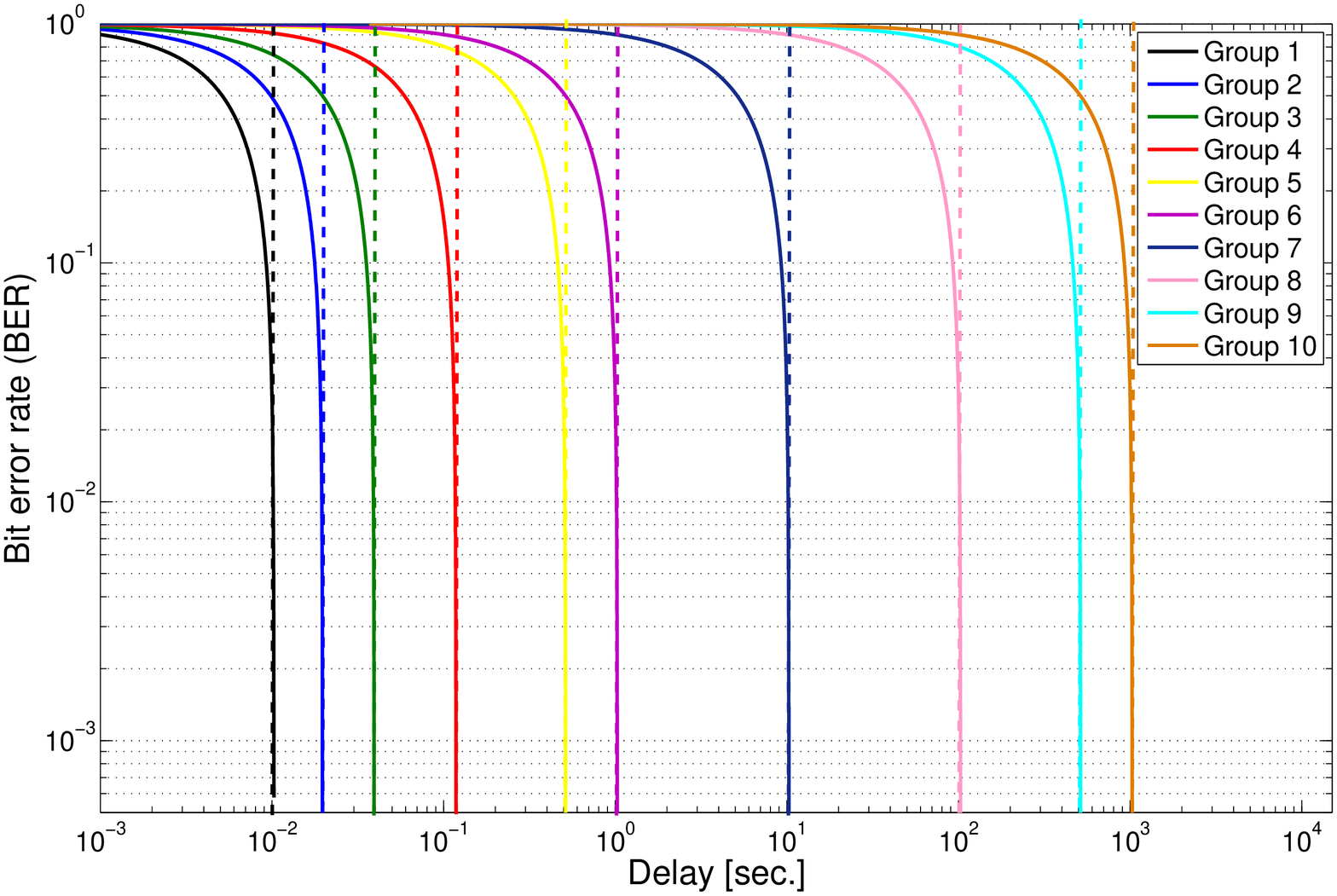}
\caption{The average bit error rate for different groups of devices classified by delay requirements. The total number of devices is $N=1000$, each group has 100 devices; the devices use the same code degree $d_c=8$, access probabilities are calculated from (\ref{pioptfor}) and $\gamma=0$ dB. Dashed lines show the delay requirement for different groups of devices.}
\label{figasym}
\end{figure}

We now consider the case that devices have different delay requirements. We assume that the payload size of each device is 1 kb and 100 MTC devices belongs to each specific delay group, as shown in Table \ref{DelayApp}. By using the optimization approach in Section IV, we can derive the optimized access probability for each group. Fig. \ref{figasym} shows the average BER versus the delay for devices in each group of MTC devices specified in Table \ref{DelayApp}, when all the devices has the same code degree $d_c=8$. As can be seen in this figure, the delay requirement for devices in various groups have been simultaneously satisfied.

\section{Conclusions}
In this paper, we proposed a novel rateless multiple access scheme to support a large number of devices in M2M communications. This is achieved by representing the multiple access process by a capacity approaching analog fountain code to minimize the access delay while achieving a higher throughput. In the proposed scheme, devices are allowed to transmit at the same time and the same channel, thus forming an equivalent analog fountain code at the base station (BS). The BS then performs joint decoding to recover all devices' messages. We further analyzed the proposed multiple access analog fountain code by using the density evolution technique and formulated an optimization problem to find the optimum access probability for each device to satisfy the delay requirements of MTC devices. Simulation results showed that the proposed scheme closely approaches the optimal common-rate performance of the multiple access channel, even for a very large number of devices. Moreover, the proposed scheme can satisfy the delay requirements for all the devices with a relatively low decoding complexity, which increases linearly with the number of devices. We also proposed a simple contention-based strategy and formulated the average overhead of the proposed scheme. Our novel random access and rateless multiple access schemes also achieved a significantly lower access delay and blocking probability compared to existing massive random access schemes for cellular-based M2M communications.
\appendices

\section{Proof of Lemma \ref{GeneralDensity}}
\label{ProfLemma1}
Let us consider the equivalent AFC code at the BS, where the equivalent generator matrix and the message vector are denoted by \textbf{G} and \textbf{b}, respectively. We first define $Y_j\triangleq\sum_{r'\in\mathcal{M}(j)\backslash r}g_{j,r'}b_{r'}$, where $\mathcal{M}(j)\backslash r$ is the set of all variable nodes connected to check node $y_{j}$ except variable node $b_r$. In \cite{guo2008multiuser}, it has been shown that the average log likelihood ratio (LLR) passed from check node $y_{j}$ to variable node $b_r$ in the $t^{th}$ iteration of the BP algorithm, $L_{j\rightarrow r}^{(t)}$, can be approximated as follows:
\begin{align}
\nonumber \mathbb{E}\left\{L_{j\rightarrow r}^{(t)}\right\}=2g_{j,r}\int_{-\infty}^{+\infty}\frac{f_1(y)}{f_0(y)}dy,
\end{align}
where $f_m(y)$ is defined as follows for $m\in\{0,1\}$:
\begin{align}
\nonumber f_m(y)=\sum_{r'\in\mathcal{M}(j)\backslash r}\frac{e^{-\frac{1}{2}\left(y-Y_j\right)^2}}{\sqrt{2\pi}}(y-Y_j)^m\prod_{r'\in\mathcal{M}(j)\backslash r}p_X(b_{r'}).
\end{align}
As shown in \cite{guo2008multiuser}, when the number of variable and check nodes go to infinity, $\int_{-\infty}^{+\infty}\frac{f_1^2(y)}{f_0(y)}dy$ tends to $\frac{1}{1+\sigma^2_Y}$, where $\sigma^2_Y$ is the variance of $Y_j$. As shown in \cite{guo2008multiuser}, in the asymptotic case the message passed from a variable to a check node is normally distributed and we can find the mean and variance of the message passed from a Type-X$_i$ variable node to a Type-\textbf{v} check node, denoted by $L_{\text{X}_i\rightarrow \textbf{v}}^{(t)}$, as follows:
\begin{align}
\nonumber
\textstyle &|\mathbb{E}\left\{L_{X_i\rightarrow \textbf{v}}^{(t)}\right\}|=\frac{2}{1+\sigma^2_{\textbf{v}}}\sum_{\ell'\in\mathcal{N}(r)\backslash \ell}g^2_{\ell',r},\\
\nonumber &\textstyle \text{var}\left\{L_{X_i\rightarrow \textbf{v}}^{(t)}\right\}=2|\mathbb{E}\left\{L_{r\rightarrow \ell}^{(t)}\right\}|,
\end{align}
where $\mathcal{N}(X_i)\backslash \ell$ is the set of all check nodes connected to a Type-X$_i$ variable node except check node $y_{\ell}$. As we assume that the number of variable and check nodes go to infinity, according to the law of large number, for large code degree $d_i$, $\sum_{\ell'\in\mathcal{N}(X_i)\backslash \ell}g^2_{\ell',X_i}$ can be approximated by $\sigma^2_wmd_i/k$, where $m$ is the number of coded symbols. As a Type-X$_i$ variable node is connected to a Type-\textbf{v} check node with probability $q_{\textbf{v}}$, then the average total LLR value for a Type-X$_i$ can be found as follow:
\begin{align}
\nonumber m^{(t)}_{i}=\gamma |h_i|^2\sigma^2_wd_{v_i}\sum_{\substack{\textbf{v}\\ v_i=1}}\frac{2}{1+\sigma^2_\textbf{v}}q_\textbf{v},
\end{align}
where $\sigma^2_{\textbf{v}}=\sum_{i=1}^N d_i\gamma |h_i|^2\sigma^2_wv_i\Omega(m_i^{(t)})$
and $\Omega(x)=\frac{1}{\sqrt{2\pi}}\int_{-\infty}^{+\infty}\left(1-\tanh(x-y\sqrt{x})\right)e^{-\frac{y^2}{2}}dy$ \cite{guo2008multiuser}. This completes the proof.

\section{Proof of Lemma \ref{ColumnLemma}}
\label{proofColumnLemma}
To generate each AFC coded symbol, $d_c$ information symbols are selected from those with minimum degrees, so after receiving $m$ coded symbols at the BS, the degree of each information symbol will be either $d_v$ and $d_v-1$, with probabilities $1+\beta-d_v$ and $d_v-\beta$, respectively. Moreover, two columns of \textbf{G} are the same if the number of their nonzero elements are the same. The probabilities that two columns of \textbf{G} have the same degree $d_v$ and $d_v-1$, are $(1+\beta-d_v)^2$ and $(d_v-\beta)^2$, respectively. Furthermore, the probability that the nonzero elements of the two columns  of the same degree are the same is $1/\dbinom{m}{a}D^a$, where $a$ is either $d_v$ or $d_v-1$. This arises from the fact that each nonzero element of each column is randomly chosen from $m$ possible positions, and the value of each nonzero element is randomly selected from $D$ possible weights. This completes the proof.
\bibliographystyle{IEEEtran}
\footnotesize
\bibliography{IEEEabrv,References}

\end{document}